\documentclass{article}

\usepackage{arxiv}

\usepackage[utf8]{inputenc} % allow utf-8 input
\usepackage[T1]{fontenc}    % use 8-bit T1 fonts
\usepackage{hyperref}       % hyperlinks
\usepackage{url}            % simple URL typesetting
\usepackage{booktabs}       % professional-quality tables
\usepackage{amsfonts}       % blackboard math symbols
\usepackage{nicefrac}       % compact symbols for 1/2, etc.
\usepackage{microtype}      % microtypography
\usepackage{cleveref}       % smart cross-referencing
\usepackage{lipsum}         % Can be removed after putting your text content
\usepackage{graphicx}
\usepackage[numbers]{natbib}
\usepackage{doi}
\usepackage{lscape} % 途中から部分的に用紙を横に使いたいしたいときはlscapeパッケージを入れて(\usepackage{lscape})、横にしたい所をlandscape環境でくくる

\title{
	Odor Maps from the LLM-derived similarity scores
}

\newif\ifuniqueAffiliation

\uniqueAffiliationtrue
\ifuniqueAffiliation % Standard variant of author block
\author{ \\
	\href{https://orcid.org/0009-0000-4254-7803}{\includegraphics[scale=0.06]{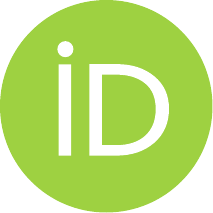}\hspace{1mm}Yuki Harada} \\% Name + ORCID
		Institute of Integrated Research, Institute of Science Tokyo, \\
		4259, Nagatsuta-cho, Midori-ku, Yokohama, Kanagawa 226-8501, JP \\
	\texttt{harada.y.ai@m.titech.ac.jp} \\
	\And
	\href{https://orcid.org/0000-0003-0620-2757}{\includegraphics[scale=0.06]{fig/orcid.pdf}\hspace{1mm}Manuel Aleixandre} \\% Name + ORCID
		Institute of Integrated Research, Institute of Science Tokyo, \\
		4259, Nagatsuta-cho, Midori-ku, Yokohama, Kanagawa 226-8501, JP \\
	\texttt{aleixandre.m.aa@m.titech.ac.jp} \\
	\And
	\href{}{\hspace{1mm}Manabu Okumura} \\% Name
		Institute of Integrated Research, Institute of Science Tokyo, \\
		4259, Nagatsuta-cho, Midori-ku, Yokohama, Kanagawa 226-8501, JP \\
	\texttt{oku@pi.titech.ac.jp} \\
	\And
	\href{https://orcid.org/0000-0002-0599-226X}{\includegraphics[scale=0.06]{fig/orcid.pdf}\hspace{1mm}Takamichi Nakamoto} \\% Name + ORCID
		Institute of Integrated Research, Institute of Science Tokyo, \\
		4259, Nagatsuta-cho, Midori-ku, Yokohama, Kanagawa 226-8501, JP \\
	\texttt{nakamoto.t.ab@m.titech.ac.jp} \\
}

\else
\usepackage{authblk}

\newbox{\orcid}\sbox{\orcid}{\includegraphics[scale=0.06]{fig/orcid.pdf}} 
\author[1]{%
	\href{https://orcid.org/0009-0000-4254-7803}{\usebox{\orcid}\hspace{1mm}Yuki Harada\thanks{\texttt{harada.y.ai@m.titech.ac.jp}}}
}
\author[1]{%
	\href{https://orcid.org/0000-0003-0620-2757}{\usebox{\orcid}\hspace{1mm}Manuel Aleixandre\thanks{\texttt{aleixandre.m.aa@m.titech.ac.jp}}}
}
\author[1]{%
	\href{https://orcid.org/}{\usebox{\orcid}\hspace{1mm}Manabu Okumura\thanks{\texttt{oku@pi.titech.ac.jp}}}
}
\author[1]{%
	\href{https://orcid.org/0000-0002-0599-226X}{\usebox{\orcid}\hspace{1mm}Takamichi Nakamoto\thanks{\texttt{nakamoto.t.ab@m.titech.ac.jp}}}
}
\affil[1]{ \\
	Institute of Integrated Research, Institute of Science Tokyo, \\
	4259, Nagatsuta-cho, Midori-ku, Yokohama, Kanagawa 226-8501, JP \\
}
\fi

\renewcommand{\shorttitle}{
	Odor Maps from the LLM-derived similarity scores
}

\hypersetup{
pdftitle={
	Odor Maps from the LLM-derived similarity scores
},
pdfsubject={
	汎用LLMを利用した匂いマップの作製
},
pdfauthor={
	Yuki Harada, 
	Manuel Aleixandre, 
	Manabu Okumura, 
	Takamichi Nakamoto
},
pdfkeywords={Odor space, Odor map, large language model (LLM), digital twin, automatic scent creation, Dravnieks, essential oil},
}

\begin{document}
\maketitle

\begin{abstract}

The application of large language models (LLMs) to OdorSpace analysis attracts growing interest. 
Recent studies have explored the comparison of sensory evaluation spaces derived from LLMs with odor character profiles in the Dravnieks' dataset. 
In this study, we calculated pairwise distances of odor descriptors using three distance measures 
and statistically compared these LLM-derived similarities with distances derived from the original data. 
Next, we extended this approach to odor names (ingredients). 
Statistical comparison revealed that LLMs can infer odor similarity to some degree, 
suggesting the potential of odor maps generated from these similarity data. 
Applying this approach, we generated an odor map of essential oils. 
It demonstrates that essential oils within the same group are closely located in the odor map, 
suggesting that the proximity in the odor map corresponds to human evaluation.

% \section*{Author summary}

% Authers' group has investigated automatic scent creation. 
% The key objective is to translate odor space into a sensory evaluational one. 
% This study demonstrates a potential to generate odor maps by LLM-derived pairwise similarities of odor names. 
% We believe LLMs hold significant promise for advancing digital scent technology and overcoming the limitations of traditional, large-scale sensory evaluation methods.

\end{abstract}

% keywords can be removed
\keywords{Odor space \and Odor map \and large language model (LLM) \and digital twin \and automatic scent creation \and Dravnieks \and essential oil}

% ======================== Main % ======================== 
\section*{Introduction}

While odor plays a significant role in health and life sciences and is deeply ingrained in human culture, 
its implementation into digital twin, Virtual Reality (VR) and related technologies lags behind those for other senses. 
Odor is represented through the molecular behavior of olfactory receptors and/or the activity of the brain. 
It is often expressed through diverse language during human sensory evaluations, and sometimes it is quantified within them.
Since odor space is very complicated, we face the challenge to predict olfactory perception. 

Several researchers have studied the prediction of olfactory perception from molecular structure parameters. \cite{keller2017predicting, khan2007predicting, gutierrez2018predicting, snitz2013predicting}
In most of cases, they predicted a set of odor discriptor for mono molecules. Recently, the olfactory perception of mixture was reported. \cite{snitz2013predicting, ravia2020measure, satarifard2025high}
However, those researches have been limited by difficultiy in obtaining large-scale perception data. 
Several researchers triied to predict olfactory perception from e-nose or mass-spectra. \cite{yokoyama1993detection, hanaki1996artificial, guo2021odrp, nozaki2016odor}
The odor descriptor in the perceptional data can be represented using Natural Language Processing (NLP). \cite{nozaki2018correction, nozaki2016odor} % [2] 
Moreover, this approach was applied to automatic scent creation. \cite{aleixandre2024automatic, aleixandre2025generative} % [2] 
However, large-scale olfactory perceptional data are essential when we study actual problem after finishing toy problem. 

Kurfal et al. investigated how AI systems can extract olfactory information based on natural language, 
paralleling human odor perception from textual data. \cite{kurfali2025representations} % [5] 
They systematically evaluated three generations of language models—static embedding models (Word2Vec, FastText), 
encoder-based models (BERT), and decoder-based LLMs (GPT-4o, Llama 3.1) —under nearly 200 training configurations. 
Their work demonstrated that LLMs can infer odor descriptors by investigating human and AI olfactory representations, 
suggesting that we can obtain large-scale perceptional data easily. 

While some analyses of 'odor space' rely on sensory evaluation, \cite{nambu2010visual, zarzo2009understanding} 
the creation of an 'odor map' remains challenging due to the substantial effort and difficulties 
in obtaining high-quality, large-scale human sensory evaluations of odors; 
however, their work offers a potential avenue for addressing this limitation. 

In this study, we focused on the similarities derived from LLM representing scent itself, alongside odor descriptors. 
We also applied this approach to create an odor map of essential oil names —a resource utilized in a previous study—
and evaluated the consistency of Multi-Dimensional Scaling (MDS) \cite{dillon1984multivariate} coordinates for ingredients within several odor groups.
No one has reported LLM-based odor map followed by dimensional analysis by MDS as far as we know.

\section*{Materials and methods}

\subsection*{Dravnieks dataset }

In 1985, Andrew Dravnieks published a semantic profiling of 180 odorants \cite{101520DS61EB}. % [6-8] 
The data used here were taken from pyrfume dataset \cite{castro2022pyrfume, githubpyrfume}. % [6-8]
A list of 146 odor descriptors (see also "Dravnieks odor descriptors.csv" 
and "dravniks stimuli for simbyllm.csv" in SI), selected from an initial list of 800, was provided to over 100 panel members from different organizations. 
These members then evaluated each descriptor's suitability on a scale of 0 to 5 for every odorant.

Fig 1(a) illustrates an example of the two-dimensional mapping generated using metric MDS. 
Pairwise cosine distances were initially calculated for the 160 odor names in the Dravnieks dataset (a total of 12,720 comparisons).
To determine pairwise distances, we employed Euclidean, Pearson correlation, and cosine distances. 
The reason for using several distance metrics is that we do not know the property of similarity obtained from LLM. 
Therefore, we should investigate which metric can be used to account for LLM similarity. 
These distances will then be compared to those from the LLM.

We utilized sklearn.manifold.MDS \cite{sklearnmds} 
and scipy.cluster.hierarchy \cite{scipyhierarchy} 
as Python libraries for MDS and hierarchical cluster analysis, respectively.

% Fig 1 two-dimensional mapping utilizing MDS of odor names, (a), (b) % 要作成・要言及 
% \begin{landscape}
\begin{figure} % {figure*}[!t]% 全幅画像挿入
    \centering
    \includegraphics[keepaspectratio, width=4.9 in ]{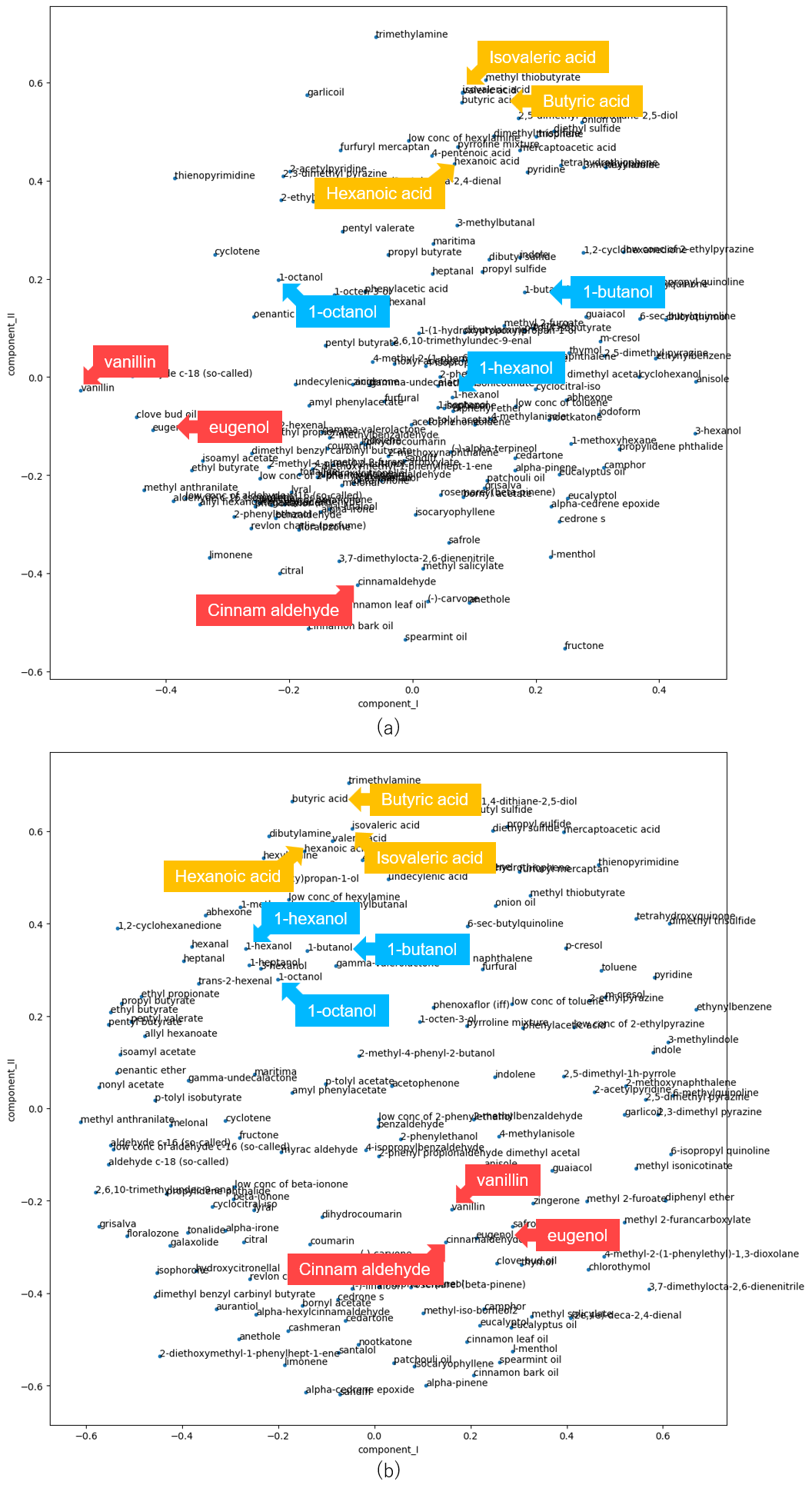}
    \caption{Two-dimensional mapping of odor names using MDS, %  MDS
    based on (a) cosine distance calculated from the Dravnieks dataset and 
    (b) similarity scores obtained from GPT-4o-mini for all pairwise for 160 odor names. }
\end{figure} % {figure*}
% \end{landscape}

\subsection*{Generating Similarity Scores with Pre-trained LLMs}

In this study, we asked the pairwise similarity to LLM. 
We implemented a prompt using OpenAI's API to generate similarity scores based on the previous work \cite{kurfali2025representations} % [5]
and the example prompts is shown in Fig 2.

% Fig 2 An example prompt to generate similarity scores; between 'cis-3-hexenol' and 'beta-ionone' 
% \begin{landscape}
\begin{figure} % {figure*}[!t]% 全幅画像挿入
    \centering
    \includegraphics[keepaspectratio, width=4.9 in ]{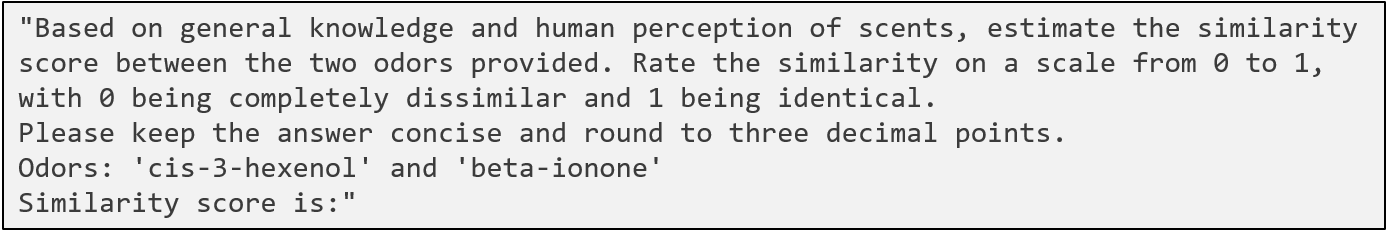}
    \caption{An example prompt to generate similarity scores; between 'cis-3-hexenol' and 'beta-ionone'  }
\end{figure} % {figure*}
% \end{landscape}

% "Based on general knowledge and human perception of scents, estimate the similarity score between the two odors provided. Rate the similarity on a scale from 0 to 1, with 0 being completely dissimilar and 1 being identical. 
% Please keep the answer concise and round to three decimal points. 
% Odors: 'cis-3-hexenol' and 'beta-ionone' 
% Similarity score is:"

The prompt was executed utilizing the Python LangChain package \cite{pypilangchain}.  
Fig 1(b) presents the resulting two-dimensional mapping of odor descriptors using MDS 
derived from 12720 pairwise similarity values obtained from GPT-4o-mini.

We queried four pre-trained LLMs: GPT-4o-mini, Gemma 3:12b, Gemma 3:4b, and Gemma 3:1b. 
These models were selected, because GPT-4o-mini was employed in previous studies, 
and the Gemma 3 models are locally executable and widely used. 
The designations 12b, 4b, and 1b represent the number of parameters in each model, where 'b' stands for billion. 
Pairwise similarity measures were obtained for 146 odor descriptors and 160 odor names from the Dravnieks evaluation set 
and will be compared in the following section.

\subsection*{Statistical comparison of pairwise metrics}

To compare the distance measures generated by LLM analysis for the Dravnieks dataset, we created a scatter plot and conducted Mantel test on it. 
the statistical tests compares two distance matrices by computing the correlation 
between the distances in the lower (or upper) triangular portions of the symmetric distance matrices. 
We utilized Python libraries; skbio.stats.distance.mantel \cite{skbiomantel} for Mantel test.

\section*{Results and Discussion}

\subsection*{Comparing Cosine Distances in Dravnieks dataset and GPT LLM-based Similarity }

Figures 1 (a) and (b) highlight alcohols with light blue, carboxylic acids with yellow, and aromatics with red, respectively. 
In both cases, chemicals with the same functional groups are located closely to each other. 
The two-dimensional MDS mapping of odor names, generated by the two metrics, 
shows that odor names within each group possess similar relative locations in both cases.

Fig 3 (a) presents a comparison of pairwise cosine distances between the 160 odor names from the Dravnieks dataset and their corresponding similarity scores from LLM, i.e., GPT-4o-mini. 
Figures 3 (b) and (c) show similarity histgrams of GPT-4o-mini scores and pairwise cosine distances in Dravnieks dataset, respectively. 
The scatter plot Fig 3 (a) shows a somewhat dispersed pattern; Mantel's statistical tests yielded a correlation coefficient of 0.332 and a p-value of 0.001. 
This statistically significant correlation suggests that the two metrics are related to each other to some extent but have limited linear relationship. 
% a medium but limited linear relationship between the two metrics. 

% Fig 3 The correlation of pairwise distances were compared on 160 odor names
% \begin{landscape}
\begin{figure} % {figure*}[!t]% 全幅画像挿入
    \centering
    \includegraphics[keepaspectratio, width=3.5 in ]{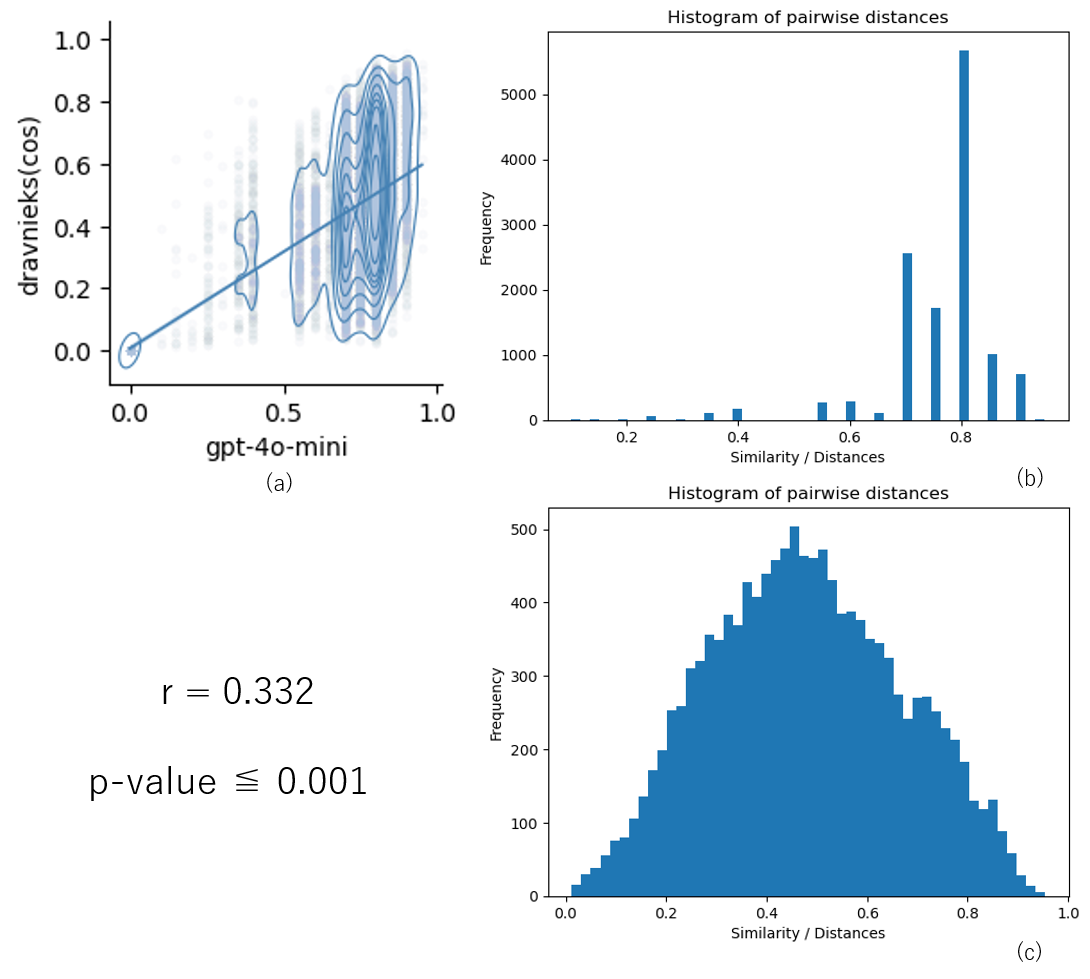}
    \caption{Comparing Dravnieks-dataset Cosine Distances and GPT-4o-mini Similarity; 
        (a) A scatter plot displays the correlation between two metrics; Contour lines show the data destribution. % the blue line shows the kernel density estimate of the data. 
        (b) A histogram presents the distribution of pairwise similarity scores generated by GPT-4o-mini. 
        (c) A histogram presents the distribution of pairwise cosine distances calculated between the 160 odor names in the Dravnieks dataset.
    }
\end{figure} % {figure*}
% \end{landscape}

% Seaborn's pairplot() function provides a quick and insightful way to visualize relationships within a dataset. 
% It generates a matrix of scatter plots, histograms, 
% and kernel density estimates (KDEs) for each pair of variables in your data. 
% This allows you to easily identify correlations, patterns, and potential outliers 
% across multiple features simultaneously. 
% It's a great tool for exploratory data analysis.

\subsection*{Comparing Distance Metrics: Odor descriptors}

We analyzed 10585 pairwise distance values on the odor descriptors in the Dravnieks dataset. 
We then compared pairwise distance across seven different metrics, 
including three distance metrics for analyzing the dataset (Euclidean, Pearson correlation, and cosine) 
and four pre-trained LLMs (GPT-4o-mini, Gemma 3:12b, Gemma 3:4b, and Gemma 3:1b) to assess pairwise similarity. 
% including three distance metrics - Euclidean distance(Eu), Pearson correlation distance(corr), and cosine distance(cos) - 
% which we employed as pairwise distance metrics for analyzing the dataset, 
% four pre-trained LLMs - GPT-4o-mini, Gemma 3:12b, Gemma 3:4b, and Gemma 3:1b - which we employed as pairwise similarity.
Fig 4 (a) presents Mantel statistical test results for the 146 odor descriptors across all metrical combinations

% Fig 4 Heatmap visualization of statistical test results across the seven metrics
% \begin{landscape}
\begin{figure} % {figure*}[!t]% 全幅画像挿入
    \centering
    \includegraphics[keepaspectratio, width=5.2 in ]{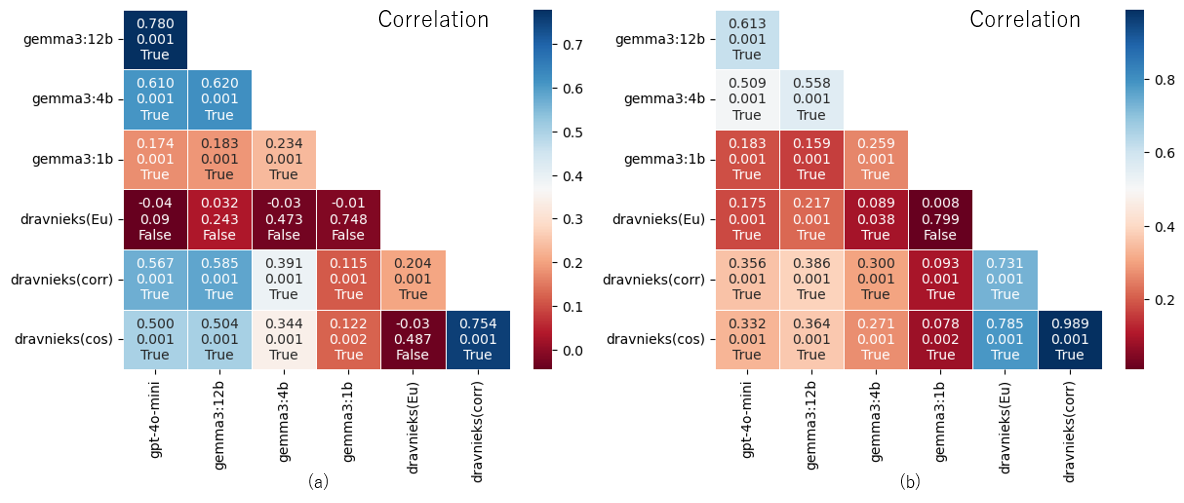}
    \caption{Heatmap visualization of Mantel statistical test results across the seven metrics for (a) the 146 odor descriptors and (b) the 160 odor names. 
        Color indicates the correlation coefficient between pairwise distances across the seven metrics 
        (four pre-trained LLMs: GPT-4o-mini, Gemma 3:12b, Gemma 3:4b, and Gemma 3:1b, 
        and three distance metrics: Euclidean distance (Eu), Pearson correlation distance (corr), and cosine distance (cos)). 
        Each cell displays a statistical test result (top), p-value (middle), and significance indicator (bottom).
        }
\end{figure} % {figure*}
% \end{landscape}

Mantel tests revealed statistically significant correlations between Dravnieks' pairwise distances and LLM-derived similarities, 
with the exception of the Euclidean pairwise distance. 
Furthermore, the tests indicated that correlation coefficients increased with LLM parameter size, 
aligning with the expectation that better reasoning capabilities lead to closer approximations of human evaluation. 
(Although the parameter count for GPT-4o-mini is not publicly available, it's believed to have reasoning capabilities similar to or exceeding Gemma 3:12b.)
The correlation between Euclidean pairwise distances and LLM-derived similarities consistently yielded small correlation coefficients and large p-values, 
indicating a lack of meaningful association. 
Given the substantial correlation between Dravnieks' data and LLM-derived similarities, 
we concluded that meaningful inferences could be drawn from GPT-4o-mini and Gemma 3:12b.

In MDS, stress is commonly used as an indicator for determining the appropriate number of dimensions in the latent space.
Fig 5(a) illustrates the relationship between Stress and the MDS dimension (n-components) based on seven different metrics .
Even when using higher numbers of dimensions (n-components), applying GPT-4o-mini and Gemma 3:12b resulted in a decrease in stress. 
This suggests that these models possess a high degree of reasoning power.

% Fig 5 Stress and the MDS dimension (n-components) based on seven different methods
% \begin{landscape}
\begin{figure} % {figure*}[!t]% 全幅画像挿入
    \centering
    \includegraphics[keepaspectratio, width=5.2 in ]{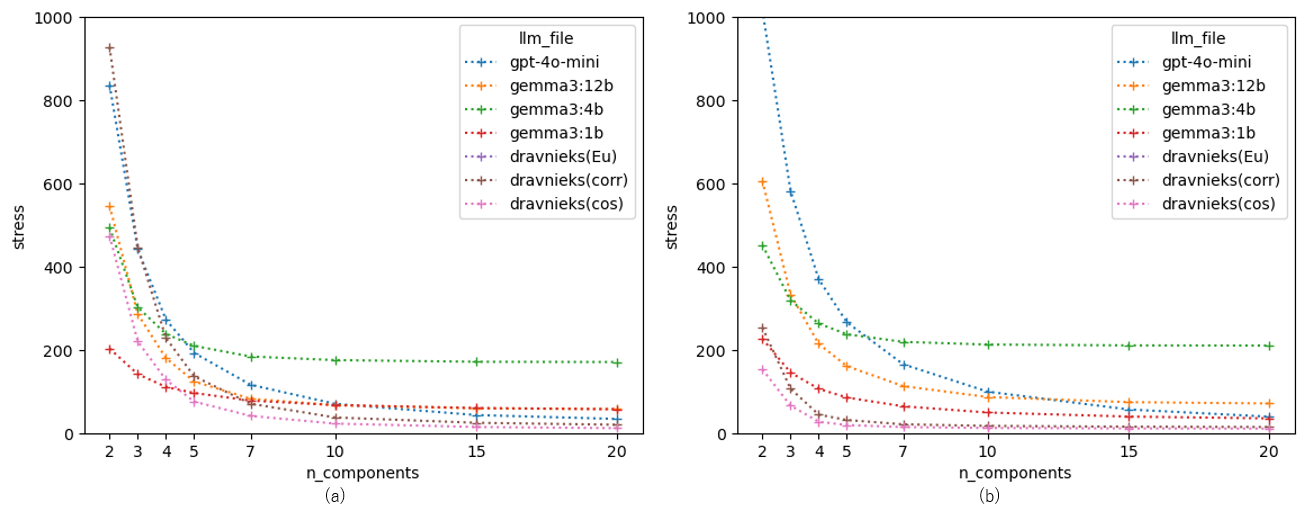}
    \caption{Stress and the MDS dimension (n-components) based on seven different methods, (a) for 146 odor descriptors, (b) for 160 odor names.}
\end{figure} % {figure*}
% \end{landscape}
% CurrentTex.SimByLLM.260319/fig/Fig6_MDSStress.png

\subsection*{Comparing Distance Metrics: Odor names}

To evaluate the consistency of the odor name analysis, 
we compared 12720 pairwise distance measures across seven different metrics. 
Fig 4 (b) presents Mantel statistical test results for the 160 odor names. 
We also found that correlation coefficients increased with LLM parameter size, 
where reasoning approches human evaluation. % parameter size 
% supporting the expectation that better reasoning aligns with human evaluations. 
The correlation between Euclidean distances and those derived from LLMs yielded smaller correlation coefficients followed by larger p-values, 
indicating a lack of meaningful association.

Fig 5 (b) illustrates the relationship between Stress and the MDS dimension for the 160 odor names. 
Applying GPT-4o-mini and Gemma 3:12b demonstrates a decrease in stress with MDS dimension, 
even it the region of high dimension. % when using larger numbers of dimensions. 
% Our analysis indicates that the reasoning capabilities of LLMs for odor names exhibited a similar trend with model parameter size as observed for odor descriptors. 
In our analysis, the similar relationship between reasoning capability and parameter size observed with odor descriptors can also be seen with odor names. 
Based on the substantial pairwise distance correlation between Dravnieks data and the relationship 
between Stress and the MDS dimension, 
we concluded that desirable odor maps can be generated using GPT-4o-mini and Gemma 3:12b.

% Fig 4(b) presents the statistical test results for the 160 odor names. 
% We also found that the correlation coefficients increased as the parameter size of the four pre-trained LLMs increases, reflecting high LLM reasoning power. 
% Except for the Euclidean pairwise distance, all other metrics were statistically significant in this instance.
% Fig 5(b) illustrates the relationship between Stress and the MDS's n-components. Also for the odor names, application of GPT-4o-mini and Gemma 3:12b demonstrates that a decrease in Stress with increased n-components even when dealing with larger n-components.
% According to our analysis, the reasoning of LLMs for odor names exhibited a similar tendency in model parameter size compared to the reasoning for odor descriptors.
% Based on the substantial pairwise distance correlation between Dravnieks data and the relationship 
% between Stress and the MDS's n-components, we concluded that desirable odor maps could be produced by GPT-4o-mini and Gemma 3:12b.

\subsection*{Odor Maps of essential oils from the LLM-derived similarity scores}

We validated that the LLM analysis has consistency with Dravnieks dataset. 
then, we applied it to 96 ingredient names from an essential oil set to create an odor map. 
We included 96 odor ingredients in the current study, 
but here focused on 75 odor names after removing extraneous information such as essential oil grade or production area.
Four types of LLMs were then used to generate 2774 pairwise similarity scores for the odor names. 
We compared the LLM-derived pairwise similarities using correlation coefficients.

% Fig 6 Result of LLM-derived similarity scores for essential oils 
% \begin{landscape}
\begin{figure} % {figure*}[!t]% 全幅画像挿入
    \centering
    \includegraphics[keepaspectratio, width=5.2 in ]{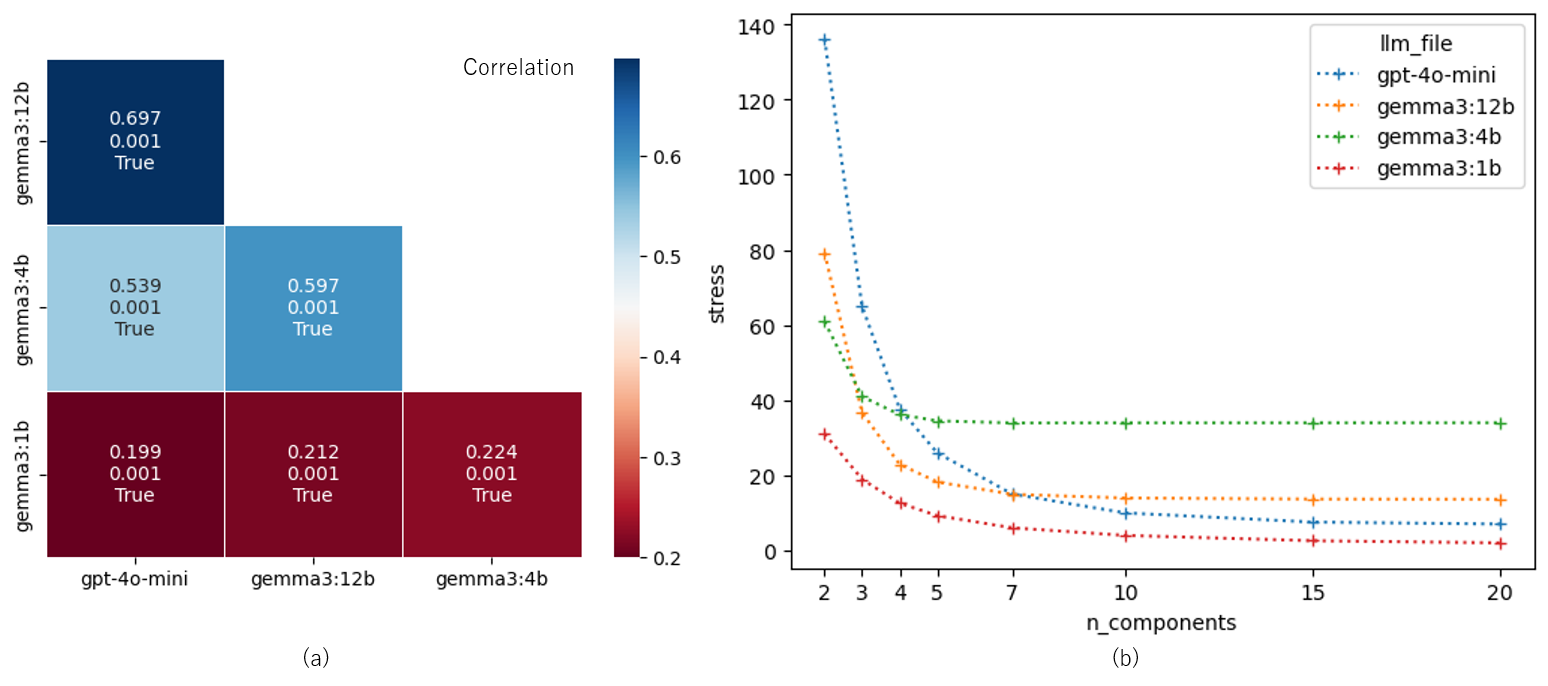}
    \caption{Result of LLM-derived similarity scores for essential oils; (a) heatmap visualization of statistical test results across four LLM metrics; 
    We calculated the correlation coefficient between the metrices derived from different LLMs. 
    values in each cell display correlation coefficients at the top, p-values in the middle, and significance levels in the bottom cells of the heatmap, (b) relationship 
    between Stress and the MDS dimension (n-components). 
    }
\end{figure} % {figure*}
% \end{landscape}
% CurrentTex.SimByLLM.260319/fig/Fig6_MDSStress.png

Fig 6 (a) presents the results of Mantel statistical tests among four LLM metric combinations. 
We obtained consistent results with the previous analysis, % for the Dravnieks dataset 
where we observed that, with statistically significance, 
% as the parameter size of the four pre-trained LLMs increased, 
the correlation coefficients also increased with the parameter size of pre-trained LLMs, 
% indicating that the LLMs' reasoning power became more close to human evaluation. 
Fig 6 (b) illustrates the relationship between Stress and the MDS dimension based on the four LLMs. 
The relationship demonstrates that GPT-4o-mini and Gemma 3:12b may exhibit high reasoning power, consistent with the Dravnieks dataset analysis; 
even with MDS dimension, applying these models resulted in a decrease in Stress with MDS dimension. 
Furthermore, the correlation coefficient also suggests that GPT-4o-mini and Gemma 3:12b yielded similar result. 
% generated desirable odor maps. 

% Fig 7 Visualization of odor space based on pairwise similarities derived from GPT-4o-mini
% \begin{landscape}
\begin{figure} % {figure*}[!t]% 全幅画像挿入
    \centering
    \includegraphics[keepaspectratio, width=5 in ]{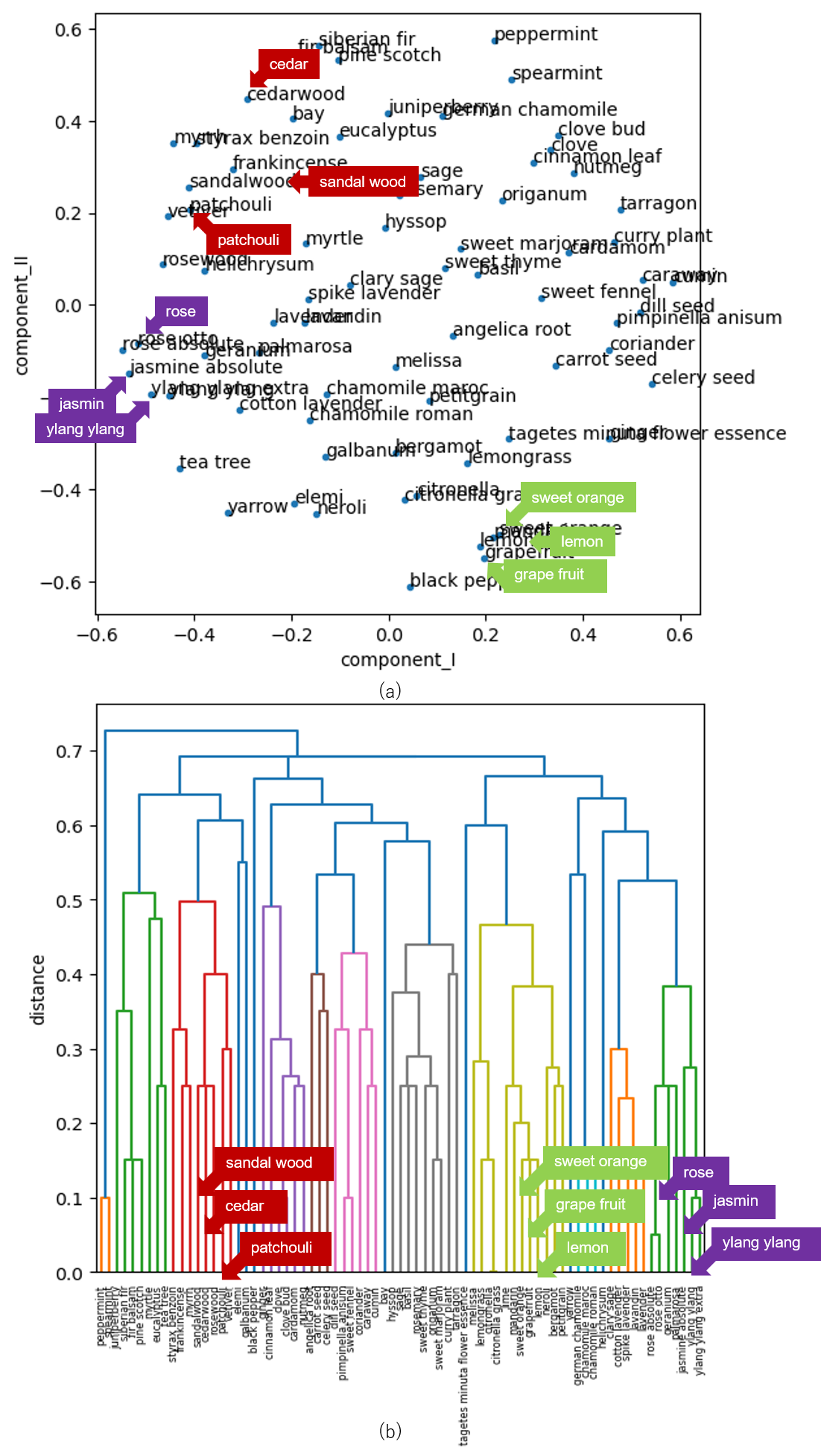}
    \caption{Visualization of odor space based on pairwise similarities derived from GPT-4o-mini. 
        (a) Odor map: 2D representation of the odor names by MDS (n-components=2). 
        (b) Dendrogram from hierarchical clustering, using the average linkage method and Euclidean distance metric.
    }
\end{figure} % {figure*}
% \end{landscape}
% CurrentTex.SimByLLM.260319/fig/Fig6_MDSStress.png
% "C:\Users\yharada\Documents\CurrentTex.SimByLLM.260319\fig\Fig_Gpt4om_EO96Ing.png"

Fig 7 (a) presents the odor map via a 2D representation of MDS based on pairwise similarities from GPT-4o-mini, 
and Fig 7 (b) dose the dendrogram from hierarchical clustering analysis. 
MDS can be executed for any dimension size, and the resulting coordinates are readily available for automated fragrance creation
—a valuable tool for aligning odor space with sensory evaluational coordinates.

To confirm the mapping and clustering of essential oils, we highlighted three groups
—those with floral scents (rose, jasmine, ylang-ylang), citrus scents (lemon, sweet orange, grapefruit), and woody scents (sandalwood, patchouli, cedarwood)—
using purple, green, and red, respectively, in Figures 6(a) and (b). The results demonstrate that 
% each group is assigned similar MDS coordinates and clusters closely.
samples in each group are located close to each other and belong to same cluster. 
Therefore, LLM-derived odor map can provides the closeness between samples since similar samples are closely located.

\section*{Conclusion}

We found that LLM-derived similarity in terms of odor name and odor descriptor matched the sensory data using Dravnieks dataset. 
We subsequently applied this approach to 
the net 75 essential oils. 
We confirmed that LLM-derived result is consistent 
% the agreement between obtaining consistent results 
with the actual sensory data i.e., the Dravnieks dataset analysis. 
Moreover, it indicates that the essential oils in the same group are closely located in the LLM-derived odor map. 
Despite the relatively low effort required, we can create a comprehensive odor map which meets
notably the need for high-quality, large-scale human sensory evaluations of odors. 
Although Dravnieks data set was freqently used in the field of olfactory perception, it appeared more than 40 years ago. 
There have been only a few work to make large-scale odor map since then because 
we have much difficulty in constructing large-scale odor map when we conduct sensory test physically. 
Large-scale odor map is very essential in the progress of digital scent technology \cite{nakamoto2025digital}. 
This odor map holds potential for automated scent creation by providing coordinate transformations of odor space in a way consistent with human sensory evaluation.

\section*{Supporting information}

% Include only the SI item label in the paragraph heading. Use the \nameref{label} command to cite SI items in the text.
\paragraph*{Dravnieks odor descriptors.csv.}
\label{behavior_1.csv.}
% \textbf{}
This file provides sensory evaluations, odor descriptors, and ingredient information from the Dravnieks dataset.

\paragraph*{dravniks stimuli for simbyllm.csv }
\label{dravniks stimuli for simbyllm.csv }
% \textbf{ }
This file contains the ingredients within the Dravnieks dataset, specifically listing odor names that were used as input prompts for the LLMs.

\paragraph*{essential oils.csv.}
\label{essential oils.csv}
% \textbf{}
This file provides a comprehensive list of 96 essential oils, with a subset of 75 used as input prompts for the LLMs.

\section*{Acknowledgments}

This work was supported by the Japan Science and Technology Agency (JST) Future Society Creation Program (Grant Number JPMJMI25H1).
The funders had no role in study design, data collection and analysis, decision to publish, or preparation of the manuscript.

\bibliographystyle{unsrtnat}
\bibliography{reference}  %%% Uncomment this line and comment out the ``thebibliography'' section below to use the external .bib file (using bibtex) .




% 	\bibitem{keshet2016prediction}
% 	Keshet, Renato, Alina Maor, and George Kour.
% 	\newblock Prediction-Based, Prioritized Market-Share Insight Extraction.
% 	\newblock In {\em Advanced Data Mining and Applications (ADMA), 2016 12th International 
%                       Conference of}, pages 81--94,2016.

% \end{thebibliography}

\end{document}